\documentclass[preprint,12pt]{elsarticle}

\usepackage{amssymb}
\journal{New Astronomy}

\begin{document}

\begin{frontmatter}

%% Title, authors and addresses

%% use the tnoteref command within \title for footnotes;
%% use the tnotetext command for the associated footnote;
%% use the fnref command within \author or \address for footnotes;
%% use the fntext command for the associated footnote;
%% use the corref command within \author for corresponding author footnotes;
%% use the cortext command for the associated footnote;
%% use the ead command for the email address,
%% and the form \ead[url] for the home page:
%%
%% \title{Title\tnoteref{label1}}
%% \tnotetext[label1]{}
%% \author{Name\corref{cor1}\fnref{label2}}
%% \ead{email address}
%% \ead[url]{home page}
%% \fntext[label2]{}
%% \cortext[cor1]{}
%% \address{Address\fnref{label3}}
%% \fntext[label3]{}

   \title{The \emph{Planck} On-Flight Forecaster (POFF)}

%% use optional labels to link authors explicitly to addresses:
%% \author[label1,label2]{<author name>}
%% \address[label1]{<address>}
%% \address[label2]{<address>}

  \author[Massardi]{Marcella Massardi}
\address[Massardi]{INAF - Osservatorio Astronomico di Padova, Vicolo dell'Osservatorio 5, I-35100 Padova, Italy}
              \ead{marcella.massardi@oapd.inaf.it}
  \author[Burigana]{Carlo Burigana}
\address[Burigana]{INAF - IASF Bologna, Via Gobetti 101, I-40129 Bologna, Italy}
             \ead{burigana@iasfbo.inaf.it}

\begin{abstract}
The \emph{Planck} On-Fligh Forecaster (POFF) is a tool to predict when a position in the sky will be within a selected angular distance from any receiver direction of the \emph{Planck} satellite according to its pre-programmed observational strategy. This tool has been developed in the framework of the \emph{Planck} LFI Core Team activities, but it is now used by the whole collaboration.

In this paper we will describe the tool and its applications to plan observations with other instruments of point sources which are expected to enhance the possibilities of scientific exploitation of the \emph{Planck} satellite data, once they will be publicly available. Collecting simultaneous multi-frequency data, like those that can be planned with the POFF, will help, on one hand, to investigate variability of point sources and, on the other, to reconstruct point source spectral energy distributions on wide frequency ranges minimizing the effects due to source variability.

POFF is a combination of IDL routines which combine the publicly available information about the \emph{Planck} scanning strategy and focal plane shape in order to identify if a given (list of) position(s) can be observable by the satellite at a given frequency and/or by selected receivers in a given time range. The output can be displayed with the desired time resolution and selecting among various sorting options.

The code is not a Planck product, but it has been validated within the \emph{Planck} LFI pipeline, looking for sources in the first satellite datasets. It will be implemented among the general tools of the LFI Data Processing Center. The code format and the large number of options make it flexible and suitable for many applications, allowing to get results quickly.

POFF is currently successfully used to plan activities within the \emph{Planck} collaboration, including observations with several ground-based facilities, and it is distributed outside it.
\end{abstract}


\end{frontmatter}

%****************************************************************************
\section{Introduction}
The ESA's \emph{Planck}\footnote{Planck \emph{(http://www.esa.int/Planck)} is a project of the European Space Agency - ESA - with instruments provided by two scientific Consortia funded by ESA member states (in particular the lead countries: France and Italy) with contributions from NASA (USA), and
telescope reflectors provided in a collaboration between ESA and a scientific Consortium led and funded by Denmark.} satellite, launched on May the 14th, in addition to improving anisotropy measurements of the Cosmic Microwave Background (CMB), is surveying the sky in nine frequency bands (33, 40, 70 GHz observed by the array of radiometers of the Low Frequency Instrument (LFI, Mandolesi et al. 1998 and 2009), and 100,143, 217, 353, 545, 857 GHz for the bolometric array of the High Frequency Instrument, (HFI, Puget et al. 1998, Lamarre et al. 2009)  with FWHM ranging from 5 to 33 arcmin). It will provide the first all-sky survey above 100 GHz (The Planck Collaboration 2006). It is expected to detect thousands of Galactic and extragalactic sources. Most of the extragalactic sources will be dusty galaxies, detected at high frequencies, but, given its sensitivity, it is expected to detect up to several hundred extragalactic radio sources (L\'{o}pez-Caniego et al. 2006, Leach et al. 2008, Massardi et al. 2009) in one or more of its lower frequency bands (30, 44, 71, 100, 143 GHz) and to measure polarisation (L\'{o}pez-Caniego et al. 2009) up to 353 GHz: this means that it will observe from 2 to 3 times more radio sources than detected in WMAP\footnote{http://lambda.gsfc.nasa.gov/product/map/current/} maps at similar frequencies (23-94 GHz, Wright et al.\ 2008, Massardi et al. 2009). A large fraction of these sources may also be detected up to 353 GHz.

Some of the radio sources will be very bright objects (prominent quasars and other flat spectrum Galactic or extragalactic sources) and will have been frequently observed in the past and well characterized. Others will be rare inverted spectrum sources with $\alpha > 0$ ($S\propto \nu^\alpha$), or other unusual sources not systematically studied or even present in low frequency catalogs. Still, others may be flaring sources with temporarily strong high-frequency emission (see Burigana 2000, Terenzi et al. 2002, 2004 for preliminary studies of \emph{Planck} capabilities to address source variability issues).

To date, the AT20G (Murphy et al. 2009, Massardi et al. 2008, 2009) provides the best ground-based sample of the high frequency sky. Given its high resolution, it mainly collects extragalactic objects. \emph{Planck} will improve on this, by providing a blind survey at even higher frequencies and at several epochs. We expect that \emph{Planck}'s all sky, high frequency and multi-frequency survey will provide a less biased sample of blazars and other extreme extragalactic sources than past ground-based works, and will consequently allow us to check features of the jet-shock paradigm introduced by most models for AGN emission. Studying the spectral energy distribution (SED) and the variability of a large sample of AGNs, including various subclasses, is a way to probe the physics of the innermost regions of the sources. With a substantial, unbiased sample, unification models of AGN (Urry \& Padovani 1995) can be tested. \emph{Planck} data together with lower frequency information may help to differentiate between unifying schemes based on Doppler and obscuration mechanism, exploiting the \emph{Planck} High Frequency Instrument to characterize the IR emission from galaxies and quasars.
Poisson fluctuations given by undetected extragalactic radio sources (i.e. sources at fluxes smaller or of the order of $200-300$ mJy) should dominate the CMB angular power spectrum at $\ell$ greater than or of the order of 1500, at \emph{Planck} LFI frequencies (Toffolatti et al. 1998, De Zotti et al. 1999, Toffolatti et al. 2005). As a consequence, it will be crucial to reconstruct the emission properties of bright extragalactic radio sources.

During outbursts, warious classes of variable Galactic radio sources show flux density levels high enough to be observed with a good S/N by \emph{Planck}.
As a remarkable example, the massive X-ray binary system Cyg X-3 is one of most active and variable Galactic radio source with quiescent periods and strong outbursts interpreted as synchrotron emission in a jet-like structure, reaching up to 20 Jy at 8.4~GHz. More than 200 Galactic X-ray binaries are known and $10 \%$  of these are radio loud (Mirabel and Rodrigues 1999). Furthermore, Luminous Blue Variable stars (van Genderen 2001), like Eta Carinae, are characterized by variability and by sudden outbursts, during which a large amount of mass is ejected from the star and the flux density can reach several Jy at centimeter wavelengths and tens of Jy in the millimeter. Finally, active binary stars, like RS CVn and Algol, have alternate periods of quiescence, with flux densities of few tens of mJy, and active periods characterized by flares, lasting several weeks, every 2-3 months, reaching several hundreds of mJy up to Jy at cm wavelengths (Umana et al. 1995).  Radio spectra show the maximum of emission at frequencies higher than 10~GHz and, during the impulsive phases, up to about 100 GHz.

In order to fully exploit and enhance the scientific impact of \emph{Planck} data, a set of supporting possible coeval observations of the same sources, with particular attention to those that show unusual spectra or flaring events, across as much of the electromagnetic spectrum as possible are being carried out with many facilities all around the world in the framework of the \emph{Planck} non-CMB science-driven activities.

\emph{Planck} observations may help identifying the properties of the SED in the high frequency radio and FIR bands, and will benefit from the ground-based observations at longer wavelength for the reconstruction of the properties of different emission components.

After the satellite launch, ESA has publicly released the information about the Planned Pointing Lists (PPL) which enclose the expected sequence of pointing positions of the satellite spin axis and the Spacecraft/Instrument Alignment Matrix (SIAM) which describes the angles of each receiver pointing direction with respect to the spin axis.

By combining these pieces of information, the POFF (\emph{Planck} On-Flight Forecaster\footnote{This tool is not an official Planck product, but it is produced on a best-effort basis by individual members of the Planck collaboration.}) code allows us to predict when a given (set of) position(s) in the sky will be within a given angular distance from any satellite receiver pointing direction.

Hence, the code can select, among a list of given positions, those that are expected to be observed within a given period by the satellite. These capabilities make it extremely useful for several purposes, and is already frequently used within the \emph{Planck} community to plan ground-based activities (and most of all observations) almost simultaneously with the satellite.

Its main use is to plan 'coeval observations' for selected samples of objects. With the term 'coeval' we mean that the positions scanned through by the satellite are observed with other instruments within a small enough period so that, if the position is associated to a source, the observation is not affected by the intrinsic variability of the source. In fact, the allowed time delay between two observations to be still considered "simultaneous" depends on the physical properties of the observed source. It is also a powerful support to the activities of LFI Data Processing Center aimed at identifying point sources in the actual satellite time ordered data, providing constrains on the observing time for given positions.

In the same spirit that led ESA to make available the PPL and SIAM information, POFF is now publicly available to plan almost simultaneous observational activities to be compared with the results of the satellite once they will be publicly released. A preliminary all-sky catalogue of compact and point sources, extracted from \emph{Planck}'s data, will be released to the scientific community approximately 10 months after the first full sky survey is finished (i.e. February 2010), in time for the first post-launch call for Herschel observing proposals.

A tool analogous to the POFF has been developed in the framework of the \emph{Planck} collaboration to track the transits of Solar System moving objects on \emph{Planck} receivers ad to plan coeval observations of those bodies detectable by the satellite (Maris and Burigana 2009).

In this paper we will describe the fundamental aspects of the \emph{Planck} scanning strategy (\S \ref{sec:Planck_ss}) and focal plane (\S \ref{sec:Planck_fp}), the condition to define when a position will be observed by a receiver (\S \ref{sec:Planck_obscond}), the structure of the POFF code (\S \ref{sec:code}) and its performances (\S \ref{sec:performances}), and some representative applications (\S \ref{sec:application}). Finally, in \S \ref{sec:conclusions} we summarize our results and illustrates the main perspective of POFF applications. In the appendices we collected a brief user manual for each input keyword and few examples of application.

%****************************************************************************
\section{The \emph{Planck} satellite pointing}

\subsection{The scanning strategy}\label{sec:Planck_ss}

{\it Planck\/} is a spinning satellite observing the sky from its Lissajous orbit around the second Lagrangian (L2) point of the Sun-Earth system periodically shifting the spin axis to remain almost anti-Sun. Thus, its receivers observe the sky by continuously scanning nearly great circles on the celestial sphere. The telescope field-of-view rotates at 1 rpm, while its spin axis regularly moves, about 2.5 arcmin/hour in ecliptic longitude, to remain almost anti-Sun. The Lissajous orbit is seen from the Earth as an about 6-month-periodic path of the spin axis pointing sequence.

A specific scanning strategy (SS) has been studied for {\it Planck\/} to minimize systematic effects and to achieve all-sky coverage for all receivers. The SS is defined by a set of relevant parameters. The main one is the angle, $\alpha$, between the spacecraft spin axis and the telescope optical axis. Given the extension of the focal plane unit (see Sect. \ref{sec:Planck_fp}), each beam centre points to its specific angle, $\alpha_{\rm r}$. The angle $\alpha$ is set to $85^\circ$ to achieve a nearly all-sky coverage even in the so-called {\it nominal} SS in which the spacecraft spin axis is kept always exactly along the antisolar direction. This choice avoids the `degenerate' case $\alpha_{\rm r} = 90^\circ$, characterized by a concentration of the crossings of scan circles only at the ecliptic poles and the consequent degradation of the quality of destriping and map-making codes (Janssen and Gulkis 1992, Wright et al. 1996, Burigana et al. 1997, Delabrouille 1998, Maino et al. 1999).

Since the {\it Planck\/} mission is designed to minimize straylight contamination from the Sun, Earth, and Moon (Burigana et al. 2001, Maffei et al. 2009, Sandri et al. 2009, Tauber et al. 2009), it is possible to introduce modulations of the spin axis from the ecliptic plane. This allows to maximize the sky coverage, keeping the solar aspect angle of the spacecraft constant for thermal stability. The adopted {\it baseline} SS\footnote{The above nominal SS is kept as backup solution in the case of a possible verification in flight of an unexpected, bad behaviour of {\it Planck\/} optics.} (Maris et al., 2006) implements this option: the spin axis performs a cycloidal modulation, i.e. a precession around a nominal antisolar direction, with a semiamplitude cone of $7.5^\circ$. In such a way all {\it Planck\/} receivers will cover the whole sky.

A cycloidal modulation with a $\sim$6 month period satisfies the mission operational constraints. This also avoids sharp gradients in the pixel hit count (Dupac and Tauber 2005). Finally, this solution spreads the crossings of scan circles in a wide region improving the quality of map-making, particularly for polarisation (Ashdown et al. 2007).

The last three SS parameters are: the sense of precession (clockwise or anticlockwise); the initial spin axis phase along the precession cone; the spacing between two consecutive spin axis repointings, chosen at $2^\prime$ to achieve four all-sky surveys with the available guaranteed number of spin axis manoeuvres.

The first all-sky survey has started on August 13th 2009. The guaranteed 15 months of observations allow the mission to independently survey the full sky at least twice with all the receivers. A request for the extension of the mission lifetime by 12 months is being processed, which may allow to complete 4 all-sky surveys.

The effective implementation of the SS of \emph{Planck} is defined by the pre-programmed directions of the spin axis that are listed in two documents (Taylor 2009):
\begin{itemize}
\item the Planned Pointing List (PPL), which contains a sequence of pointings for the \emph{Planck}'s spin axis covering about one month period of the survey.
\item the Long Term Preprogrammed Pointing List (LTPPL), which provides about one year of planned pointings and is based on the best survey information available at the time.
\end{itemize}
Their periodic updates can be downloaded from the webpage\\
http:\//www.rssd.esa.int/index.php?project=Planck\&page=Pointing.
 \begin{figure}
   \centering
   \includegraphics[width=9cm]{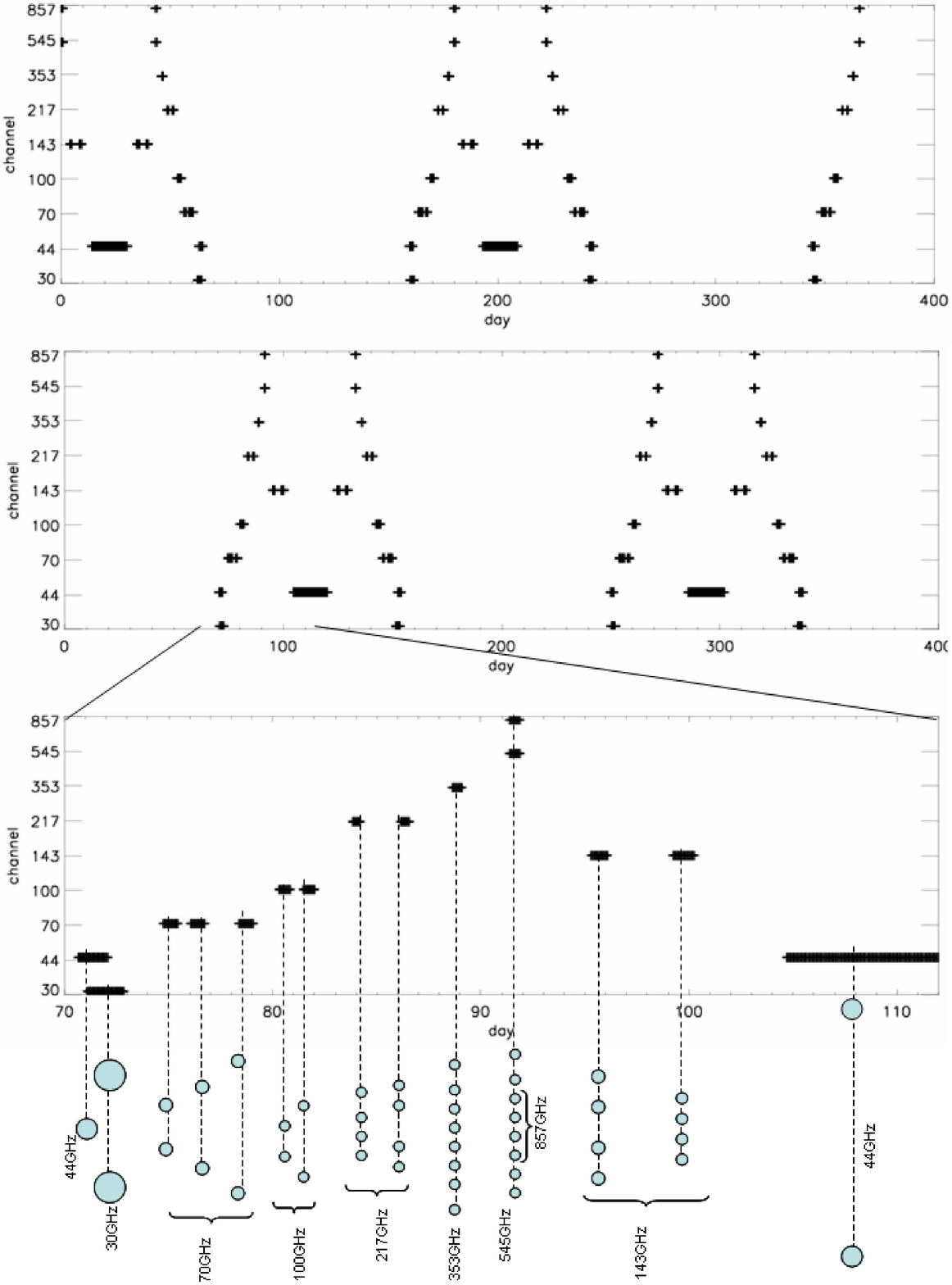}
   \caption{In the upper and middle panel there is a plot of the POFF output as a function of time and \emph{Planck} frequency channels for the South and North Ecliptic Pole positions, respectively. Day `0' is the 18th of August, first day listed in the PPL file used for this run of the code. The lower panel is a zoom in the upper one for the first period of observability of the North Ecliptic Pole position. The scheme of the \emph{Planck} focal plane has been plotted for comparison: the sequence of frequencies of observations corresponds to the sequence with which aligned beams scan through the source. Note that the distances between beams is not on scale for plotting purposes.}
              \label{fig:polo}%
    \end{figure}

\subsection{The focal plane} \label{sec:Planck_fp}

The focal plane is oriented at an 85$^\circ$ angle with respect to the spin axis, so that the scanning tracks pass close to the Ecliptic poles, but it does not cross perfectly on them.
The \emph{Planck} telescope focuses radiation from the sky onto its focal plane, which is shared by the Low and High Frequency Instruments. The field of view of the ensemble of the \emph{Planck} receivers has a global angular size of 7.5$^\circ$.

Typically, horns at the same frequency, or subsets of them, observe the same scan circle for a given spin axis direction (see Fig.\ref{fig:polo}), improving the sensitivity. These groups of aligned receivers at the same frequency are typically close each other, with the exception of the 44 GHz channel that is characterized by two aligned receivers on one side of the focal plane and a third one on the opposite side (Tauber et al. 2009, Sandri et al. 2009).

The Spacecraft Instrument Alignment Matrix (SIAM, Taylor 2008), which gives the pointing direction and orientation of each receiver of the two \emph{Planck} instruments with respect to the spacecraft functional reference frame comes originally from ground calibration. When bright sources (and, in particular, external planets) will transit on the \emph{Planck} field of view, they will be used to better characterize the optical properties of each \emph{Planck} beam, i.e. its shape, orientation, and, in particular, the pointing direction of its centre with respect to the spacecraft functional reference frame (Burigana et al. 2001). When the estimation of beam centre direction changes significantly with respect to the previous one, the SIAM file will be updated. It can be retrieved on the same webpage as the PPLs.

The first element of the part of the SIAM corresponding to a given receiver provides the cosine of the angle, $\alpha_{\rm r}$, between the direction of the considered receiver and the direction of the spacecraft spin axis. This is the quantity crucial for the POFF.

\subsection{Observability conditions} \label{sec:Planck_obscond}

For a given spin axis direction, each receiver describes a circle in the sky defined by the corresponding angle $\alpha_{\rm r}$ from the spin axis.
In reality, the spin axis direction is not exactly constant between two consecutive repointings (a pointing period), because of the complexity of spacecraft dynamics. The spread of the spin axis directions within a pointing period is of the order of $1$ arcmin or less.

We adopted the nominal beam size (i.e. 1 $\sigma$) given by the \emph{Planck} Blue Book (The Planck Collaboration 2006) to have correct indications of the observability. Then, we used as FWHM the values 33, 27, 13, 9.5, 7.1 arcmin respectively for 30, 44, 70, 100, and 143 and 5 arcmin for all the channels between 217 and 857 GHz. The size of the beam, assumed Gaussian, is $\sigma=\mathrm{FWHM}/\sqrt{8 {\rm ln}{2}}$.
For a given spin axis direction, a direction in the sky is considered observed by a given receiver if its angular distance from the circle in the sky described by that receiver is less than a certain value related to the beam size (and possibly including the spread of the spin axis directions during a pointing period).

The \emph{Planck} SS implies that positions close to the ecliptic poles will be observable throughout a long period (as long as 3 months, see Fig.\ref{fig:polo}), while objects close to the ecliptic equator will remain in the region observed by the \emph{Planck} focal plane at most for about one week during each survey (Burigana et al. 2000).

In the context of \emph{Planck} activities, a code, to a certain extent similar to the POFF, has been developed to identify the presence of signals from Solar System moving objects in the \emph{Planck} time ordered data and to plan multifrequency observations of these bodies almost coevally with \emph{Planck} (Maris and Burigana 2009). In that case the most difficult part is the requirement for a link to precise celestial mechanics codes.
Obviously, the POFF does not need to cope with celestial mechanics, because the directions of relevant sources can be considered fixed in the sky for any practical application. However, it needs a more complicate interface to cope with all the possible uses, as described in the following sections.

%****************************************************************************
\section{The POFF code}

\subsection{Structure of the code}\label{sec:code}
   \begin{figure}
   \includegraphics[width=9cm]{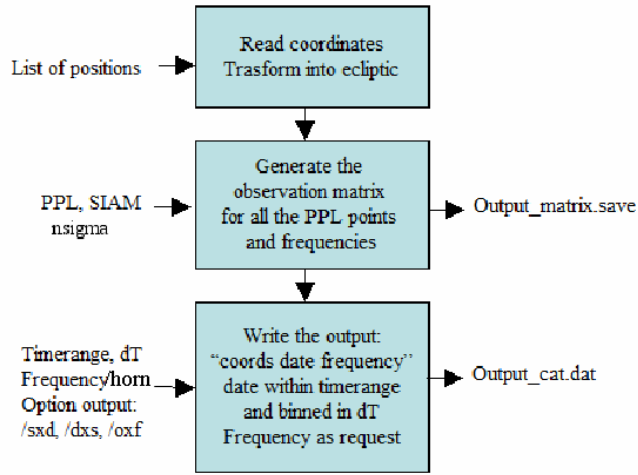}
   \caption{I/O flow chart of the POFF code.}
   \label{fig:scheme}
   \end{figure}
As described in the previous sections, the POFF code identifies the region of the sky swept by each receiver by combining the spin axis position at every pointing and the angle of each receiver with respect to the spin axis.

The code is a combination of several IDL routines, and the input and output options are passed through a set of mandatory or optional keywords. Fig. \ref{fig:scheme} shows the I/O flow chart of the code: the user provides the list of coordinates, POFF transforms them into Ecliptic, if necessary, and passes them to a routine that reads the PPL and SIAM files. The code verifies for each input pair of coordinates and for each of the \emph{Planck} receivers if it falls within the observable region during a whole rotation of the satellite around its spin axis for each PPL position.

Let's enter into details.

The input positions can be listed in a text file (the keyword that passes this information is \texttt{filein}) or passed through the command line (\texttt{posarray}). Options allow to indicate if the positions are in Ecliptic (\texttt{/ecl}) or Galactic (\texttt{/gal}) J2000 coordinate systems (default is in the Equatorial system). Coordinates should be in decimal degrees, as default, but an option (\texttt{/hours}) allows to introduce the first coordinate in decimal hours. The code performs most of its activities in Ecliptic coordinates and for this reason, in case of any different format of the input a transformation of coordinates is performed exploiting the EULER IDL routine. In the output file the coordinates are written both in the input system and in the Ecliptic one used for the calculations.

It is also possible to indicate a column of the input file which lists the identification for each coordinate pair (\texttt{colid}): default identification for each position is an integer equal to the sequential position of the pair of coordinates in the input file or in the position array.

A position is observable by a receiver if it is within, by default, one beam unit strip during the full rotation of the satellite around its spin axis. A multiplying factor can be given as input (through the keyword \texttt{nsigma}) to modify the size of the observable region.

The finest time resolution for which we could define the observability of an object is equal to the minimum separation between two consecutive positions in the PPL file ($\sim$1 hour).

PPL and SIAM information are read in the files that can be downloaded from ESA web site (their location on the user machine can be passed through the keywords \texttt{pplfile} and \texttt{siamfile}).

The code identifies receivers that observe at the same frequencies and retrieves the observability information for each frequency channel. An option (\texttt{horn}) allows to retrieve the observability information for all the receivers or only a selected subsample of them.

Once all the input have been collected, the code calculates for each given position its distance from each horn direction at any PPL pointing: if this is smaller than the size of the observable region then the source is flagged as observable at the time coinciding with the PPL pointing for the considered receiver.

In practice, the code generates an `observation matrix' of dimension [number of sources, number of frequency channels, number of PPL positions] (note that if the information on single receivers is requested the size is [number of sources, number of receivers, number of PPL positions]): an element [s, f, p] of this matrix is equal to 1 if the source s is observable in the channel (receiver) f at the PPL position p. The PPL position corresponds to an exact epoch during the satellite scan: that allows one to date the observability. The matrix, the epochs associated with each PPL position, and the list of source identifications are saved into a `\texttt{outpath}.save' file (where \texttt{outpath} is a mandatory keyword). In this way it is possible to restore for any future use the observation matrix (via the \texttt{/mat} keyword) for the same input positions list, PPL/SIAM files and option frequency/receiver, with the possibility to change the output requests.

Finally, the matrix and the epochs associated with each PPL position are given as input to a routine that produces the output catalogue of observations (stored in the file `\texttt{outpath}\_outcat.dat') according to the user requests. The user can select the frequency channels (with the \texttt{freq} keyword, default is all the channels), or the receivers, define a time range (via the \texttt{timerange} option, default is all the time for which the PPL are available) for the output epoch and a binning of the time to be printed. The range of time for which the output has to be printed is provided in the format [YYYYMMDDHH$_\mathrm{begin}$, YYYYMMDDHH$_\mathrm{end}$] and it can be binned on the wished bin size ranging from $\sim$1h (i.e. the time between a repositioning of the spin axis) to 183 days (i.e. the time necessary to complete a full survey of the sky) with the \texttt{dT} option (default is 1 day).

According to the aim for which the code is used, different sets of output may be necessary. The user can decide to list the output by epoch of observations (\texttt{/sxd}) or following the sorting order of the input positions as they have been provided at the beginning (default). In both the case the results may be ordered according to the frequency (\texttt{/oxf}).

See the Appendices for a short manual of each keyword and some specific I/O examples.

\subsection{The POFF performances}\label{sec:performances}

The POFF is distributed for free on the webpage\\ http://web.oapd.inaf.it/rstools/POFF\footnote{Under construction} in IDL binary format (i.e. a `poff.sav' file). To be executed it needs to be restored within an IDL environment (version 7.0 or following) and the calling sequence, with all the options, can be passed through the IDL command line or introduced in IDL scripts.

The actual executed pointings may differ from the pointings in the PPL in case of unpredictable repointing activities of the satellite in which case the discrepancies between the predictions and the observed pointings shouldn't be larger than the satellite focal plane (i.e. the prediction is expected to be always correct within a week time); the corresponding accuracy of the LTPPL is lower because of the longer look-forward timescale. Hence, if the user requires that the prediction is correct on shorter time scales, the short-term PPL, which are periodically updated, should be used.

The code has been validated by comparing its prediction with the First Light Survey observations of Planck (i.e. the first few days of science-driven observations of the satellite) and finding the observed sources within at most few hours from the predictions. It should be stressed that an uncertainty in the prediction which may be significant for some applications cannot be avoided, because of unpredictable adjustments in the implementation of each PPL, as well as short-term changes in the pointing sequences.

The SIAM is generated by the LFI and HFI consortia and contains the best knowledge of the focal plane properties which will evolve through the lifetime of \emph{Planck}. Currently, the relative placement of the beams is known to a fraction of the smallest beam size, i.e. of order of 2 arcmin. Their absolute location with respect to the spin axis will change during the mission and are currently provided with an accuracy of some arcminutes.
The code reads local text files containing the PPL and SIAM information in the format in which they are released by ESA.

Aiming at producing a flexible tool, we have allowed, as mentioned in the previous section, a broad set of possible input and output options.
  \begin{figure}
   \centering
   \includegraphics[width=7cm, angle=90]{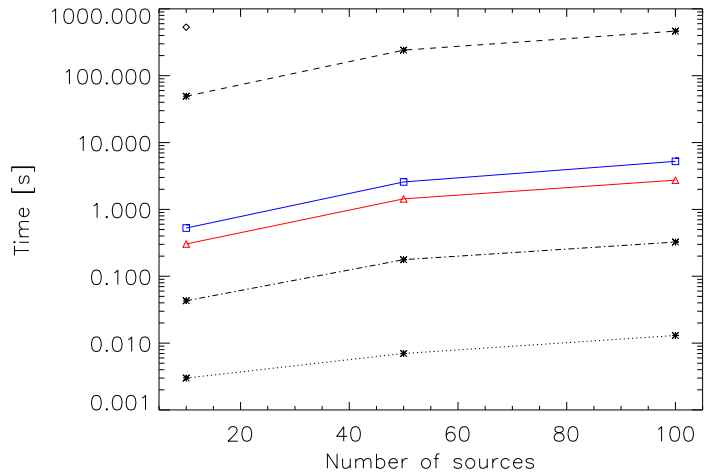}
   \caption{Summary of CPU time (in s) for the POFF runs as a function of number of positions in the input list necessary to perform different operations within the same run: input-reading (dotted line), matrix-generation (dashed line), output-writing (dot-dashed line) and total running time.
   The runs have been performed on a common laptop. The examples refers to lists of equatorial coordinates in a text file without identification column, output time range of 1 month with default binning. Different symbols corresponds to different I/O conditions: output requested for one frequency channel reading input from a text file (asterisks, note that in this case the total time almost overlaps with the matrix-generation phase time and hence we haven't plotted it), and reading the input from the `.save' matrix generated in a previous run (triangles); reading the input from the `.save' matrix but writing output for all the 9 \emph{Planck} frequency channels (squares); reading input from a text file and writing the output for 3 receiver of the same frequency (diamond).}
   \label{fig:time}
   \end{figure}

Figure \ref{fig:time} shows the time (in seconds) which is necessary to complete the various activities for different lists of positions and different input/output options. Of course the time requested for any operation increases (even if sub-linearly) with the number of sources. If we divide the activities in the three operative phases shown in Fig. \ref{fig:scheme} (i.e. trivially input reading, generation of the matrix, output writing) we can notice that the generation of the matrix is the most time-consuming operation. The input-reading phase is longer in the case in which the \texttt{/mat} option is set because it incorporates both the input-reading and matrix-generation phases, but is far shorter than the sum of these two phases in the same input conditions. The output-writing phase depends significantly on the positions (i.e. as we will see below, positions in different regions of the sky have a different probability to be observable), the time range, the time binning and the number of frequency channels to be considered.

%****************************************************************************
\section{Discussion: applications of the POFF}\label{sec:application}

\subsection{The POFF output for remarkable sources}
POFF allows to take advance of knowing the positions of the satellite in order to plan ground- or satellite-based activities.
It has a broad set of applications within the \emph{Planck} community. One of the most important possibilities is to predict when a relevant object position will be observed by the satellite. Furthermore, this is crucial to coordinate the activities when a source that can be used as a reference to set the calibration quality will be observable. To assess the quality of satellite measurement, in fact, it is useful to identify well-known sources which observation may help to check the satellite calibration both in total intensity and in polarisation.

For non-varying objects the time delays between the satellite and any other observation is not an issue: other instrument observations can help to confirm and assess the quality of the satellite measured flux densities, if their frequency is close to the satellite frequency channels, or, otherwise, to extend the measured spectral behaviour. In case of variable sources, also other observations should be coordinated in order to provide further possible checks of the satellite outcome quality.
   \begin{figure}
   \centering
   \includegraphics[height=9cm]{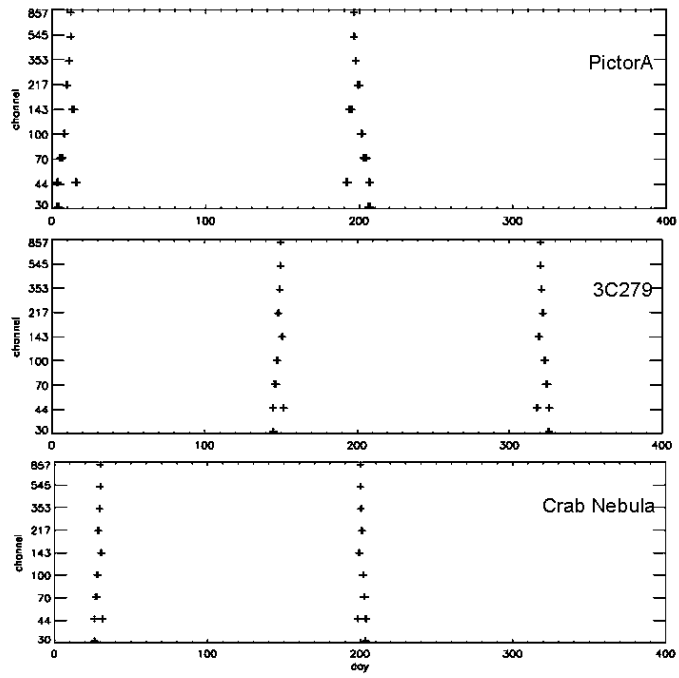}
   \caption{Plot of the POFF output for the first two Planck surveys as a function of time and frequency for Pictor A, 3C279 and the Crab Nebula. Day `0' is the 18th of August.}
              \label{fig:crab_3c279}%
    \end{figure}
Fig.\ref{fig:crab_3c279} shows the observability predictions per day and frequency channels for Pictor A, 3C279, and for the Crab Nebula which are considered suitable reference sources for polarisation calibration. While the Crab Nebula is not varying, 3C279 is a very bright blazar for which, to be suitable in this role, it will be necessary to have ground based simultaneous observations in total intensity and polarisation at some \emph{Planck} frequencies. Note that while the Crab and 3C279 lay close to the Ecliptic plane and, for this reason, they can be observed for less than a week during a whole survey, Pictor A is closer to the Ecliptic pole, so that it might be observed by the satellite, at different frequencies, for a slightly longer period.

\subsection{The POFF applications to observational campaigns}
Massardi et al. (2009) have demonstrated that almost simultaneous ground-based data are useful to fine tune the point source detection methods applied to satellite maps like the WMAP or the Planck ones. A set of samples selected at different flux density levels may help to define the reliability and completeness of the detections and to quantify the error on flux density estimation. The comparison with ground based data is extremely useful for point source analysis in the case of polarisation, since detection techniques applied to WMAP and to simulations of \emph{Planck} polarisation maps, so far, gave poor results for source polarised flux density estimation (Lopez-Caniego et al. 2009). Unfortunately, the lack of knowledge of complete samples over large area of the sky at \emph{Planck} frequencies makes the effort difficult: extrapolations of flux densities measured in other spectral bands do not guarantee the completeness of an extragalactic source sample at the \emph{Planck} frequencies, because of the complicate spectral behaviour of extragalactic the sources in this band.

In the Southern hemisphere the Australia Telescope 20 GHz (AT20G) survey helps to select samples over a large area of the sky, complete down to $<100$ mJy with measurements in total intensity and polarisation. Hence, thanks to the POFF code it is possible to plan the observations of suitable samples of AT20G sources at various frequencies during the \emph{Planck} mission. By observing close to the \emph{Planck} observations we can minimize the effects of variability.

Most of the very bright compact AGNs that will constitute the bright radio frequency population show variability over various timescales (from the intraday variable objects to flares that may last for years). The AT20G observations have demonstrated that (Sadler et al. 2006) the median variability index over a 1 year time range at 20 GHz was 6.9\% and only $\sim$5\% sources vary by more than 30\% in flux density.

Different classes of Galactic objects may show variability on different timescales. As mentioned in the Introduction, while for LVBs and X-ray binaries outbursts may show up on irregular time scales, the active binary stars alternate quiescent phases and flaring periods on time scales from weeks to months. Considering such variability time ranges is crucial in the organization of useful coeval observational activities.

Hence, a tool like POFF simplifies the planning of the ground-based observations performed during the \emph{Planck} mission can be compared with the \emph{Planck} data minimizing the variability effects, as necessary to complete the spectral energy distribution observed by the satellite with the information from other spectral bands. Meanwhile, by repeating the observations throughout the satellite surveys it will be possible to investigate how the source variability behaves with frequency.

\begin{figure}
   \centering
   \includegraphics[width=8cm, angle=90]{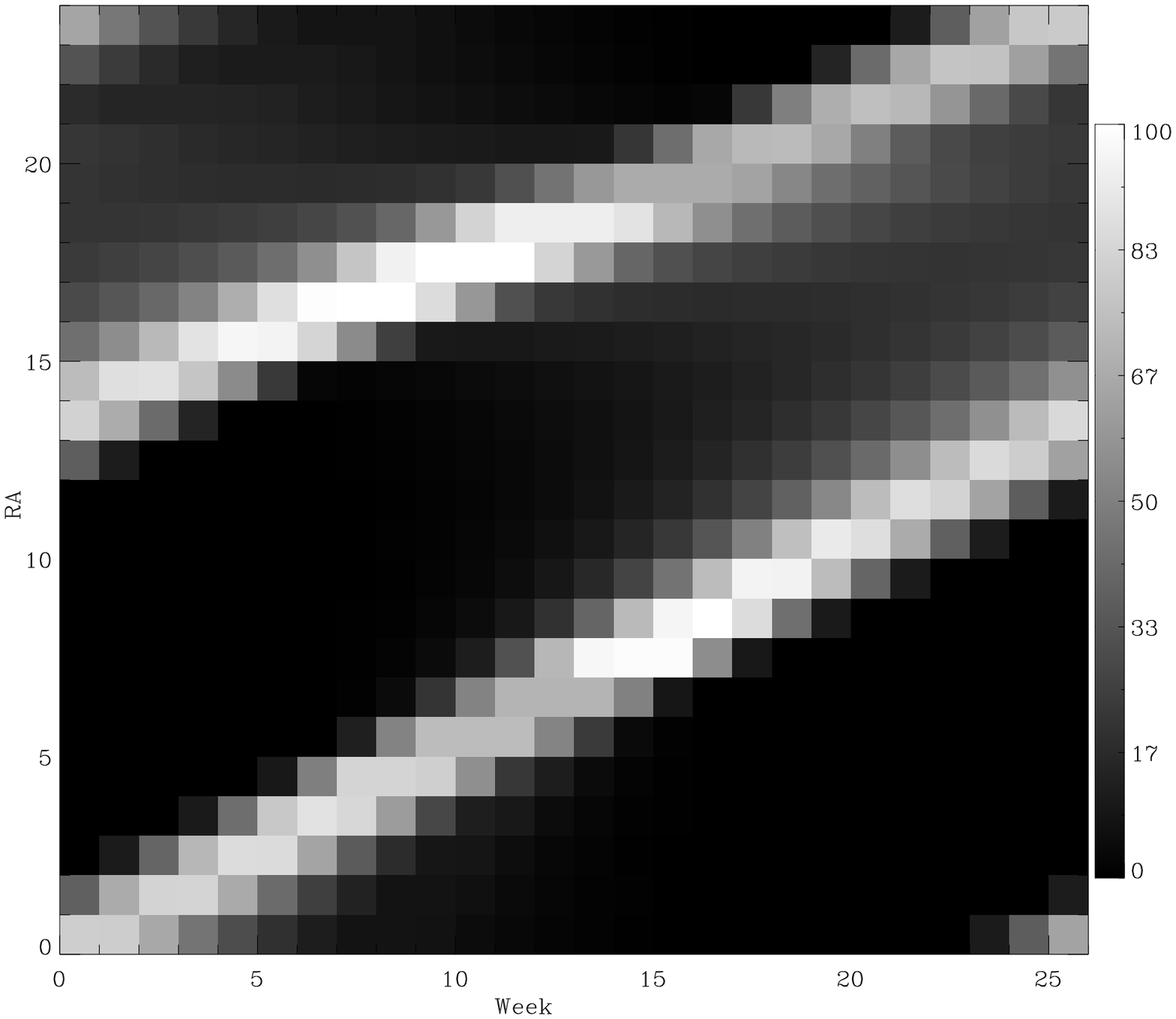}
   \includegraphics[width=8cm, angle=90]{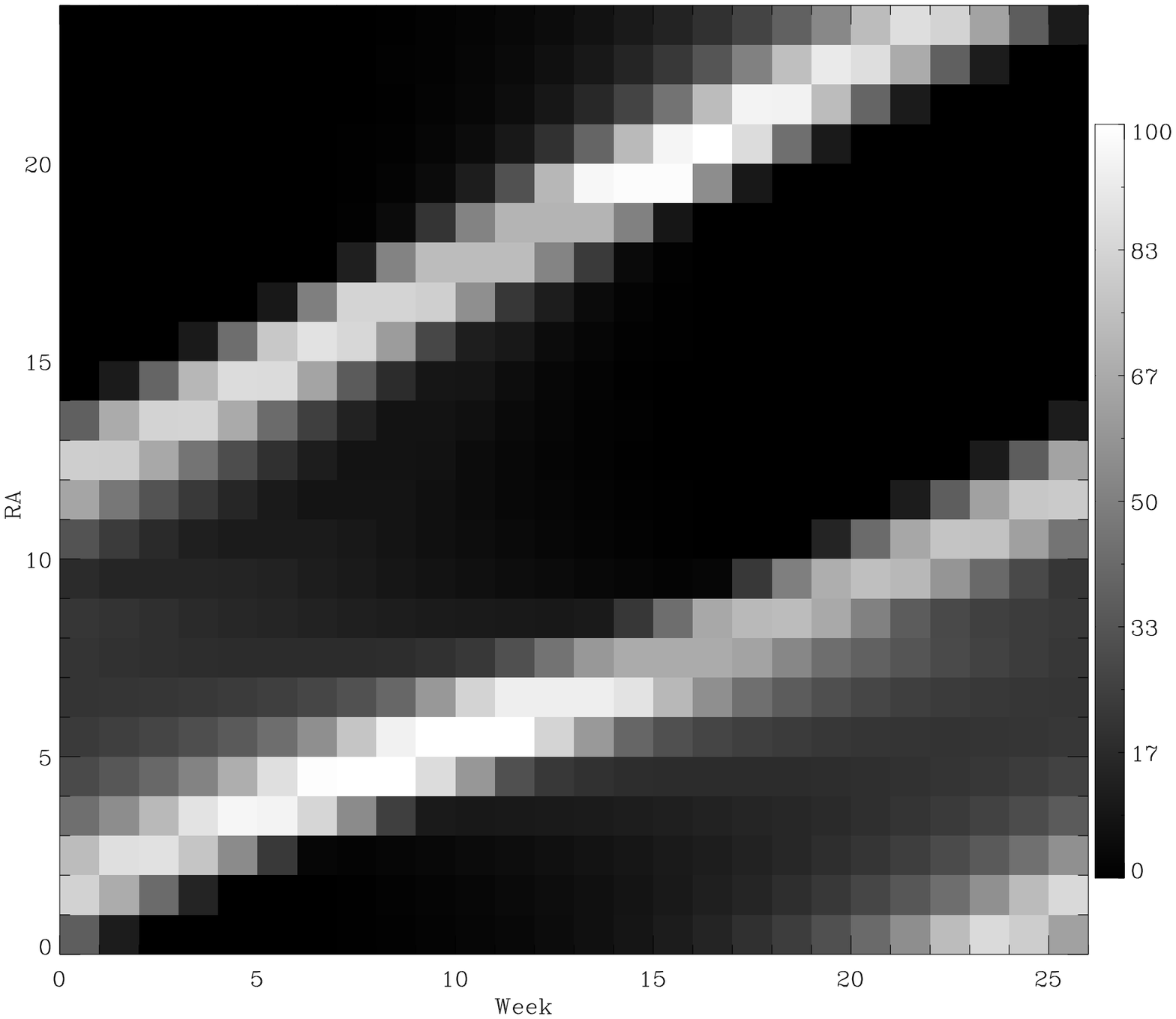}
   \caption{Fraction of right ascension region observable for each week of a \emph{Planck} survey in the Northern (upper panel) and in the Southern (lower panel) Equatorial hemispheres. The first day in of the first \emph{Planck} survey is the 18th of August. The schemes are the same for any following surveys (unless any major pointing correction is applied).}
              \label{fig:weekplan}%
    \end{figure}

In Fig. \ref{fig:weekplan} we show the fraction of right ascension regions that will be observed by the satellite in the Southern and Northern hemispheres according to the current LTPPL and SIAM files, obtained by applying POFF to a grid of positions in the sky. Since most telescopes plan their observations according to the Local Sidereal Time, that at a first approximation indicates the right ascension of the sources passing through the meridian, these plots give an idea of the LST ranges that every week may be best suited for coeval observations, throughout the whole mission (the pattern repeats at every survey).

Several projects use POFF to plan observations with ground-based telescopes to observe selected samples of objects. Among the others, we list the following on-going projects as examples of the possible exploitations of the POFF code to coordinate observational campaigns:
\begin{itemize}
\item a large sample (more than 300 objects) of extragalactic sources are being observed with the Australia Telescope Compact Array in the frequency range between 4.5 and 40 GHz; the sample includes some complete samples selected within the AT20G survey and a selection of sources that showed variability in the past and some well-known blazars; the selected sources are observed simultaneously to the \emph{Planck} observations;
\item a sample of blazars observed with the Very Large Array (VLA) in the range 1.4-43 GHz, extends the previous project to the Northern Celestial hemisphere;
\item the Atacama Pathfinder EXperiment (APEX) is being used to extend the data collection to the sub-mm band for a sample of well-known blazars also included in the ATCA sample;
\item the Noto radiotelescope will be used to observe at 43 GHz a complete sample of extragalactic radiosources in the region surrounding the North Ecliptic Pole selected in the CRATES catalogue (Healey et al. 2008), which includes low-radio-frequency selected flat spectrum sources (i.e. more probably variable objects) completed in a small region with steep spectrum sources selected combining GB6 (Gregory et al. 1996) and NVSS (Condon et al. 1998) information; furthermore, a sample of few selected Galactic objects (i.e. microquasars and binary active systems) will be observed simultaneously with the satellite and followed-up for several epochs.
\end{itemize}

%****************************************************************************
\section{Conclusions} \label{sec:conclusions}
The \emph{Planck} On-Fligh Forecaster (POFF) is a tool to predict when a position in the sky will be within one (by default) beam unit from any selected receiver of the \emph{Planck} satellite according to the preprogrammed \emph{Planck} scanning strategy. This tool has been developed in the framework of the \emph{Planck} LFI Core Team activities and is going to be implemented among the LFI Data Processing Center general tools.
The code provides the dates in which the satellite will observe each position given in an input list for each frequency channel and/or satellite receiver, so that it is possible to plan a priori the observations of known objects.

The POFF code has demonstrated to be a precious tool in planning ground-based observations simultaneous with the satellite. Many activities are on-going exploiting several facilities of the two hemispheres, in order to get the advantage of the pointing knowledge on one hand to minimize the effects of source variability, and to investigate the source variability on the other, once the ground-based observational results will be compared with \emph{Planck} data.

The code is available in IDL pseudo-executable binary format on the dedicated webpage http://web.oapd.inaf.it/rstools/POFF\footnote{Under construction}.

\section{Acknowledgements}
We gratefully thanks Andrea Zacchei and Paolo Natoli for the help in retrieving documentation and distributing the code.
We thank Gianfranco De Zotti and Grazia Umana for numberless discussions about extragalactic and Galactic source properties.
We thank Anne Lahtemaaki, Bruce Partridge, Merja Tornikoski, and Corrado Trigilio, PIs of the cited ground-based observational campaigns, that allowed us to mention their proposals in the present paper.
We thank Jan Tauber and Reno Mandolesi for the useful comments given us for this paper.

Partial financial support for this research has been provided by the Italian ASI (contracts \emph{Planck} LFI Activity of Phase E2 and I/016/07/0 `COFIS').

%****************************************************************************
\appendix
\section{POFF user manual}
The calling sequence is the following:\\
POFF, outpath, pplfile$=$pplfile, siamfile$=$siamfile, filein$=$filein, mat$=$mat, colpos$=$colpos, colid$=$colid, posarray$=$posarray, gal$=$gal, ecl$=$ecl, hour$=$hour, timerange$=$timerange, dt$=$dt, freq$=$freq, sxd$=$sxd, oxf$=$oxf, horn$=$horn,\\ nsigma$=$nsigma, man$=$man

The input are:
\begin{itemize}
\item the (list of) position(s) for which the user wants to predict the observability with \emph{Planck} according to the nominal scanning strategy defined by the PPL and SIAM.
\item the PPL and SIAM files in the format that can be downloaded from the \emph{Planck} ESA website.
\end{itemize}

The code generates:
\begin{itemize}
\item a text file that lists the observed positions according to the sorting options that have been set.
\item a `.save' file that collects the information about the observability of the given list of positions at all the frequencies or for all the receivers in the time period for which the PPL files lists the pointings. It allows to make faster future runs of the program by modifying only the output options.
\end{itemize}

The keywords available are the following:
\begin{description}
\item[\texttt{outpath}] path to generate the output files
\item[\texttt{pplfile}] the PPL file downloadable from the Planck livelink or from the ESA website. No default.
\item[\texttt{siamfile}] the SIAM file downloadable from the ESA website. No default.
\item[\texttt{filein}] name of the input file. It can be a text file (with read permissions) that lists the positions and/or the IDs of the sources or a `.save' file previously generated by this program, if the \texttt{/mat} keyword is set. Coordinates should be in degrees if the keyword \texttt{/hour} is not set. Blank spaces are considered column separators: hence no blank space is allowed within a value to be read (i.e. names like 'PKS 1921-293' are not allowed, 'PKS1921-293' is a valid name, but ra and dec columns must be separated by a blank space).
\item[\texttt{mat}] set this keyword if the filein is a `.save' file generated by the program in a previous run. This keyword is useless if the filein keyword is not set.
\item[\texttt{colpos}] 2-d array `[a,b]' where `a' is the index of the column of filein that lists the first coordinate of the positions (i.e. RA or longitude), and `b' is the index of the column of filein that lists the second coordinate of the positions (i.e. declination or latitude). The index of the first column is `1'. Default is `[1, 2]'. It is useless if \texttt{/mat} is set.
\item[\texttt{colid}] the index of the column of filein that lists the identifications of the positions. It is useless if the \texttt{/mat} keyword is set. If not set, the sequential number of the source in the filein (or in the posarray) is used as identification.
\item[\texttt{posarray}] array of coordinates in the format `[[a$_0$,b$_0$],[a$_1$,b$_1$],...]' where `a$_i$' is the i-th first coordinate (i.e. RA or longitude), and `b$_i$' is the i-th second coordinate of the positions (i.e. declination or latitude). If also \texttt{filein} is set the posarray will be concatenated at the end of the array of positions read in filein. Coordinates should be in degrees if keyword /\texttt{hour} is not set. Notice that one among \texttt{posarray} or \texttt{filein} `must' be set. It is useless if \texttt{/mat} is set. No identification can be given to these positions. If there is no filein, or if identifications are set in the filein, the sequential order of each position in the given posarray is used as its identification, otherwise, if filein is set but no identification is given, the sequential order in the concatenated list is used for every position in the list.
\item[\texttt{gal}] set this keyword if the positions are given in J2000 Galactic coordinates. If unset J2000 Equatorial coordinates are assumed.
\item[\texttt{ecl}] set this keyword if the positions are given in J2000 Ecliptic coordinates. If unset J2000 Equatorial coordinates are assumed.
\item[\texttt{hour}] set this keyword if the first coordinate of the positions (i.e. RA or longitude) is given in hour. If unset the first coordinate is assumed to be in degrees.
\item[\texttt{timerange}] 2-d array in the format `[a,b]' where `a' is the beginning and `b' the end of the time range for which the program verifies the observability of the positions. Required format is `[YYYYMMDDHH, YYYYMMDDHH]'. Default is `[2009010100, 2015010100]'.
\item[\texttt{dt}] time bin (in days) required for the output. The timerange is binned in intervals of width equal to dT and for each position observability is verified within each bin. Default is 1 (i.e. one day). Range of allowance is `[0.041, 183]' (i.e. about [1 hour, 6 months]).
\item[\texttt{freq}] n-d array listing the frequency channels required for output (dimension up to n=9). Allowed values are 30, 44, 70, 100, 143, 217, 353, 545, 857. Default is all the channels.
\item[\texttt{horn}] this keyword allows to retrieve the information about single receivers. If set it will select all the horns observing at the frequencies given through the keyword \texttt{freq} (i.e. if \texttt{freq} is not set it means "all the receivers"). It can be set to a n-d array (dimension up to n=47) listing the name(s) of the requested horn(s). Allowed values are: `lfi27', `lfi28', `lfi24', `lfi25', `lfi26', `lfi18', `lfi19', `lfi20', `lfi21', `lfi22', `lfi23', `100\_1', `100\_2', `100\_3', `100\_4', `143\_1', `143\_2', `143\_3', `143\_4', `143\_5', `143\_6', `143\_7', `143\_8', `217\_1', `217\_2', `217\_3', `217\_4', `217\_5', `217\_6', `217\_7', `217\_8', `353\_1', `353\_2', `353\_3', `353\_4', `353\_5', `353\_6', `353\_7', `353\_8', `545\_1', `545\_2', `545\_3', `545\_4', `857\_1', `857\_2', `857\_3', `857\_4'. If \texttt{freq} is not set and \texttt{horn} is set to an array, only the selected horns will be considered for the output. Default is no information on single receivers. If this keyword is set the observational matrix stored in the `.save' output file will have dimension [number of source, 47, number of PPL positions], so it can be restored only to provide information on single receivers (but they can be different from the case when the matrix has been generated).
\item[\texttt{nsigma}] multiplying factor to modify the size of the observable region for each horn. Default is 1, which means that the observable size is 1 beamwidth.
\item[\texttt{sxd}] if set, sorts the observations by dT and then according to the sequence of input list of positions. If not set, default is to list all the observations for each position sorted by dT.
\item[\texttt{oxf}] if set, sorts the observations for each position (or those for each dT if sxd is set) by frequency as first criterion.
\item[\texttt{man}] if this keyword is set, the calling sequence is written to help the user. Execution is halted.

\end{description}
\section{I/O examples}

a) Verifying observability for PicA, 3C279, and the Crab nebula in August 2009 at 30 GHz. Save the results in files with path=`picA\_3c279\_crab'. Use the PPL stored in the file `aug.PPL' and the siamfile in `0007.SIAM'

Calling sequence:\\
POFF, `picA\_3c279\_crab', pplfile$=$`aug.PPL', siamfile$=$`0007.SIAM', posarray$=[[79.957045,  -45.779019], [194.0465271, -5.7893119],
[83.633212, 22.014460]]$, timerange$=[2009080100, 2009083000]$, freq$=[30]$

Output:\\
\begin{small}
\% Compiled module: EULER.\\
\% SAVE: Portable (XDR) SAVE/RESTORE file.\\
\% SAVE: Saved variable: MATRIX.\\
\% SAVE: Saved variable: JULIAN\_DATES.\\
\% SAVE: Saved variable: LON.\\
\% SAVE: Saved variable: LAT.\\
\% SAVE: Saved variable: NAMES.\\
\% Compiled module: DAYCNV.\\
End
\end{small}
In the picA\_3c279\_crab\_output.dat file:\\
\begin{small}
\#Observations of each position ordered by date and by frequency\\
\#RA[hr]      Declination  Ecl\_lon      Ecl\_lat         YYYYMMDDHH.h  Freq   IDs\\
    5.3304698  -45.7790184   70.5812427  -68.5423592   2009082512.0    30   0\\
\end{small}
The run takes about 16 seconds on a common laptop.

b) Restore the picA\_3c279\_crab.save file generated in the previous example to verify the observability of the same sources at all the frequencies in September 2009

Calling sequence:\\
POFF, 'picA\_3c279\_crab\_sep', filein$=$'picA\_3c279\_crab.save', /mat,\\ timerange$=$[2009090100, 2009093000]

Output:\\
\begin{small}
Loading the existing matrix.\\
\% RESTORE: Portable (XDR) SAVE/RESTORE file.\\
\% RESTORE: Save file written by user@PC-USER, Mon Oct 05 23:50:22 2009.\\
\% RESTORE: IDL version 5.5 (Win32, x86).\\
\% RESTORE: Restored variable: MATRIX.\\
\% RESTORE: Restored variable: JULIAN\_DATES.\\
\% RESTORE: Restored variable: LON.\\
\% RESTORE: Restored variable: LAT.\\
\% RESTORE: Restored variable: NAMES.\\
End
\end{small}
In the picA\_3c279\_crab\_sep\_output.dat file:\\
\begin{small}
\#Observations of each position ordered by date and by frequency\\
\#RA[hr]      Declination  Ecl\_lon      Ecl\_lat         YYYYMMDDHH.h  Freq   IDs\\
    5.3304698  -45.7790184   70.5812427  -68.5423592   2009090112.0   353   0\\
    5.3304698  -45.7790184   70.5812427  -68.5423592   2009090212.0   545   0\\
    5.3304698  -45.7790184   70.5812427  -68.5423592   2009090212.0   857   0\\
    5.3304698  -45.7790184   70.5812427  -68.5423592   2009090312.0   143   0\\
    5.3304698  -45.7790184   70.5812427  -68.5423592   2009090412.0   143   0\\
    5.3304698  -45.7790184   70.5812427  -68.5423592   2009090512.0   143   0\\
    5.3304698  -45.7790184   70.5812427  -68.5423592   2009090612.0    44   0\\
    5.5755473   22.0144596   84.0976187   -1.2944772   2009091612.0    30   2\\
    5.5755473   22.0144596   84.0976187   -1.2944772   2009091612.0    44   2\\
    5.5755473   22.0144596   84.0976187   -1.2944772   2009091712.0    30   2\\
    5.5755473   22.0144596   84.0976187   -1.2944772   2009091712.0    70   2\\
    5.5755473   22.0144596   84.0976187   -1.2944772   2009091812.0    70   2\\
    5.5755473   22.0144596   84.0976187   -1.2944772   2009091812.0   100   2\\
    5.5755473   22.0144596   84.0976187   -1.2944772   2009091912.0   217   2\\
    5.5755473   22.0144596   84.0976187   -1.2944772   2009092012.0   143   2\\
    5.5755473   22.0144596   84.0976187   -1.2944772   2009092012.0   353   2\\
    5.5755473   22.0144596   84.0976187   -1.2944772   2009092012.0   545   2\\
    5.5755473   22.0144596   84.0976187   -1.2944772   2009092012.0   857   2\\
    5.5755473   22.0144596   84.0976187   -1.2944772   2009092112.0    44   2\\
    5.5755473   22.0144596   84.0976187   -1.2944772   2009092112.0   143   2\\
    5.5755473   22.0144596   84.0976187   -1.2944772   2009092212.0    44   2\\
\end{small}
The run takes about 0.12 seconds on a common laptop.

\bibliographystyle{elsarticle-num}

\end{document}